\documentclass[11pt]{article}

\usepackage[margin=1.1in]{geometry}
\usepackage{amsmath}
\usepackage{amssymb}
\usepackage{booktabs}
\usepackage{enumitem}
\usepackage[hidelinks]{hyperref}
\usepackage{lmodern}
\usepackage[T1]{fontenc}
\usepackage{microtype}
\usepackage{xspace}

\newcommand{\sys}{Quipu\xspace}
\newcommand{\op}{\textsc{shape-aware}\xspace}
\newcommand{\ctx}{\textsc{context-merge}\xspace}
\newcommand{\base}{G_{b}}
\newcommand{\ours}{G_{o}}
\newcommand{\theirs}{G_{t}}

\title{\textbf{Shape-Aware Three-Way Merge for RDF Knowledge Graphs}\\
\large The schema defines the conflict: an implementation and evaluation}
\author{Steve Brown\thanks{\textsc{orcid}
\href{https://orcid.org/0009-0009-1720-1785}{\nolinkurl{0009-0009-1720-1785}}.
With implementation and drafting assistance from Claude (Anthropic) and Codex (OpenAI).
\textbf{Artifact availability:} the merge operator, the divergence benchmark
(\texttt{benchmark/mergebench/}) and the recorded-divergence replay corpus
(\texttt{benchmark/replay/}) behind every number reported here are in the
development repository, \texttt{github.com/intigraph/quipu}. Each result table
names the command that regenerates it.}}
\date{11 November 2026}

\begin{document}
\maketitle

\begin{abstract}
Merging RDF serializations creates ordering conflicts and misses semantic ones.
We describe and evaluate a three-way merge in which the schema defines a
conflict: set algebra combines canonical triples, and a slot is contended only
where a SHACL shape bounds its cardinality and both sides' changes cannot stand
within that bound. Unbounded predicates union; contended slots go to a person. The same shapes graph validates
the result.

We do not claim the first three-way merge for RDF. Our contribution is an
implementation in a production multi-agent store and an evaluation against
a reimplementation of Quit Store's Context Merge overlap rule, which it extends
rather than replaces, and six other strategies.

Across five synthetic seeds, the shape-aware operator raised $33$ conflicts,
all true positives as judged by the same shapes that define them, with no
triples lost or fabricated and no SHACL violations admitted. The context-overlap baseline, counted per slot, detected the same $33$ conflicts but
asked $845$ questions. A replay built from recorded production incidents covers $105$ human alias
repairs ($93$ after excluding $12$ chained pairs) and $939$ duplicate-value
incidents. It reproduces a triple-level limit: different names for one entity
remain invisible ($0$ of $93$ alias repairs detected, matching $0$ of $21$
synthetic cases).

In a study of evolving shape contracts, merging shape
declarations first handles the tested cardinality changes, while datatype
evolution exposes a detector blind spot caught by post-merge validation.
An oracle that reads no shapes agrees with the shipped operator on the facts
and history of all $23$ reachable single-slot cases. Under wrong shapes, a missing bound loses data silently
and a spurious one only adds questions. The supported claim is fewer
human decisions with nothing silently lost, for cardinality conflicts in one
deployment, not general merge correctness.
\end{abstract}

\section{Introduction}

A knowledge graph that more than one writer can edit will diverge, and something
has to put it back together. The obvious move is to reuse the tooling that
already solves this for source code: keep the graph in a file, keep the file in
git, and let a line merge reconcile two versions.

That move fails in a specific and instructive way. A line merge understands a
document as an ordered sequence of lines, and a graph is neither ordered nor a
sequence. Two writers who add unrelated facts about the same entity produce
adjacent edits and a textual conflict. Two writers who re-serialize the same
unchanged graph produce a diff of everything. Meanwhile a writer who sets a
second value on a property the ontology says holds exactly one --- the change
that genuinely cannot be reconciled without a decision --- produces a clean,
conflict-free merge, because the two values live on different lines and nothing
in the file says they compete.

The failure is not that line merge is bad at graphs. It is that conflict is a
\emph{semantic} property of the data model, and a line merge is being asked to
infer it from a syntactic one. Our position is that an RDF merge should take
conflict from the place the data model already defines it: the schema.

\paragraph{This paper.} We describe an operator with two halves. Triples are
canonicalized and merged as sets, so serialization order cannot manufacture a
conflict. A SHACL shapes graph then decides which slots the set algebra is
allowed to settle on its own: where a shape bounds a slot's cardinality and
the two sides' changes cannot both stand within that bound, the operator refuses and records a
decision; everywhere else, it unions. Because the shapes are already present ---
they are what validates writes into the store --- the merge policy is not a
second configuration to keep in step with the ontology. It \emph{is} the
ontology.

\paragraph{What we are not claiming.} Three-way merge over canonicalized RDF is
not new, and we searched before writing. Quit Store~\cite{arndt2019quit} places
RDF under git with merge strategies, including a Context Merge that identifies
conflicts by node overlap and refers them to a person; TerminusDB offers diff and patch over a versioned graph; RDF
Patch and LD Patch standardize change description; RDFC-1.0 gives canonical
labeling for graphs with blank nodes, which we build on rather than beside. To
our knowledge --- and scoped to that search --- what has not been reported is an
evaluation of \emph{SHACL-derived} conflict semantics against those neighbors
in a deployed system. That evaluation is the contribution, and the framing of
the operator as an extension of Context Merge rather than a replacement for it
is deliberate.

\paragraph{Contributions.}
\begin{enumerate}[leftmargin=*,itemsep=2pt]
  \item A three-way merge operator over canonical triples whose conflict
        predicate is derived from SHACL cardinality, with post-merge validation
        against the same shapes, implemented in a production store
        (\S\ref{sec:operator}, \S\ref{sec:impl}).
  \item A divergence benchmark with an oracle recomputable by a reader from
        published data alone, scoring eight strategies under one accounting
        contract, and a result that contradicts one of our own hypotheses
        (\S\ref{sec:synthetic}).
  \item A replay of recorded multi-agent divergence from a production graph
        against the shipped operator, which reproduces the synthetic arm's
        blind spot on real data and quantifies the benefit on a class that
        production surfaced nothing of (\S\ref{sec:replay}).
  \item A study of evolving shape contracts that merges the declarations
        before the data, which handles the tested cardinality changes and
        shows that a datatype change escapes the conflict detector and is
        caught only by post-merge validation (\S\ref{sec:schema}).
  \item Two checks that do not read the shapes: a use-case oracle run against
        the shipped operator, and a re-judging of the benchmark under
        deliberately wrong shapes, which shows that a missing bound loses
        data silently while a spurious one only costs questions
        (\S\ref{sec:oracle}, \S\ref{sec:synthetic}).
  \item A stated limit --- alias divergence is invisible to triple-level merge
        --- reported as a measured false-negative column rather than excluded
        from the benchmark, with a separate entity-resolution stage tested
        against it (\S\ref{sec:limits}).
\end{enumerate}

\section{Background}
\label{sec:background}

\paragraph{The unit of change.} An RDF graph is a set of triples. Two
serializations of one graph may differ in every byte --- Turtle groups by
subject, N-Triples does not, and neither fixes an order within a group. A diff
over serialized text therefore reports differences that do not exist in the
data. Taking the triple as the unit of change removes that entire category:
a three-way merge becomes set algebra over $(\base, \ours, \theirs)$, and no
result depends on how any side chose to write its file.

Blank nodes are the exception, since they have no stable name across
serializations. RDFC-1.0 (formerly URDNA2015) assigns canonical labels and makes
graph-level equality decidable. Our operator sits on top of that substrate; the
benchmark in \S\ref{sec:synthetic} uses ground graphs only, so it does not
exercise blank-node canonicalization and is not evidence about it.

\paragraph{Where conflict comes from.} Set algebra alone answers only part of
the question. If both sides add a triple, the union holds both. That is correct
for a person's email addresses and wrong for their date of birth, and nothing in
the triples distinguishes the two cases. The distinction is in the schema: a
SHACL shape~\cite{knublauch2017shacl} with \texttt{sh:maxCount 1} on a property
path states that the slot
holds at most one value. A merge that unions into such a slot has not resolved
the divergence; it has produced a graph that violates the very constraint the
store enforces on ordinary writes.

This gives a conflict predicate with three properties we care about. It is
\emph{declared}, not inferred from edit patterns. It is \emph{already present},
because the shapes are what the store validates against. And it is
\emph{auditable}, because the same shapes graph can be run over the merge output
--- so the definition of a conflict and the test for a bad resolution are one
artifact rather than two that can drift apart.

\paragraph{Shares.} The operator reconciles \emph{share bundles}: a directory
holding a canonical N-Triples export, the shapes that govern it, and a manifest
carrying a content hash of the graph and the identifier of the share it was
derived from. The parent pointer is what supplies $\base$, so fork-point
anchoring is by content hash rather than by wall-clock time or by a
centrally-issued sequence number. Two peers who have never spoken can therefore
establish a common ancestor from the bundles alone.

\section{The operator}
\label{sec:operator}

Let $\base$, $\ours$ and $\theirs$ be sets of canonical triples, and let $S$ be a
SHACL shapes graph. A \emph{slot} is a pair $(s, p)$ of subject and predicate;
we write $G[s,p] = \{\, o : (s,p,o) \in G \,\}$ for the values a graph carries in
a slot. The slot is the unit of conflict because it is the unit
\texttt{sh:maxCount} is declared over.

\paragraph{Additions and removals.} Each side's change relative to the base is
$A_x = G_x \setminus \base$ and $R_x = \base \setminus G_x$ for $x \in
\{o, t\}$. These are exact: no heuristic decides whether a triple was
``modified'' rather than removed and re-added, because at triple granularity
there is no such distinction to make.

\paragraph{Bounded slots.} Let $\mathit{max}_S(s,p)$ be the maximum
cardinality $S$ declares for $p$ on $s$, or $\infty$ where none is declared. A
slot is a candidate only when both sides changed it and disagree:
\[
  \ours[s,p] \neq \theirs[s,p]
  \;\wedge\;
  \ours[s,p] \neq \base[s,p]
  \;\wedge\;
  \theirs[s,p] \neq \base[s,p].
\]
For a candidate, let $M[s,p]$ be the slot's three-way set merge: the base
values neither side removed, together with the values either side added. The
slot is \emph{contended} when
\[
  |M[s,p]| > \mathit{max}_S(s,p)
\]
or when $\mathit{max}_S(s,p) = 1$, the base held a value, and one side emptied
the slot while the other replaced its value. The predicate therefore yields
three classes, which the evaluation reports separately. \emph{Functional
divergence}: a slot bounded at one receives two different values.
\emph{Cardinality overflow}: a slot bounded at $n > 1$ would exceed $n$ once
both sides' additions are united. \emph{Delete/modify}: one side deleted a
functional value that the other replaced; the set merge is within the bound, so
the count alone would let the replacement silently override the deletion. An
unbounded slot is never contended.

Both sides must have moved. If only one side changed a bounded slot, the
other side's value is still the base value, there is no disagreement, and the
change applies --- a merge that stopped to ask about it would be charging a
person for a decision nobody needs to make.

\paragraph{The merge.} If any slot is contended, the operator returns
\texttt{conflicts} and writes nothing; each contended slot is reported with its
base, ours and theirs values, and the declared cardinality that made it a
conflict. Otherwise the result is
\[
  G_m \;=\; \bigl( \ours \cup \theirs \bigr) \setminus \bigl( R_o \cup R_t \bigr),
\]
a retraction on either side being honoured against the other side's retained
triples. The transaction records \emph{both} parent share identifiers, so the
merge is itself provenance: the lineage of a merged graph is inspectable in the
same way a merge commit's is.

\paragraph{Two properties worth stating.} The operator never lands a value for a
slot it flagged --- a conflicted merge writes nothing at all --- so it cannot
score well by both refusing a decision and quietly resolving it. And because
$G_m$ is defined by set operations over canonical triples, it is commutative in
$\ours$ and $\theirs$ and independent of serialization.

\paragraph{Post-merge validation.} $G_m$ is validated against $S$. This is not
belt-and-braces: it is the falsifiable form of the design claim. If conflict
semantics really are derivable from cardinality, then a merge that consulted the
shapes should not be able to produce a graph that breaks a \texttt{sh:maxCount}
bound. The claim is scoped to that constraint: datatype, class and minimum-count
constraints are outside the predicate, and \S\ref{sec:schema} reports a datatype
violation that only this validation step catches. Any
violation in $G_m$ is corruption the operator admitted without charging anyone
for it, and \S\ref{sec:synthetic} reports that column for every strategy
precisely so the claim can fail.

\section{Implementation}
\label{sec:impl}

The operator ships in \sys, a governed bitemporal RDF store, as
\texttt{share\_merge} (approximately 480 lines of Rust). It exposes two entry
points. \texttt{status} loads an incoming bundle, resolves the common ancestor
through the manifest's parent pointer, and reports what each side added and
removed together with the contended slots --- without writing anything, so a
peer can be inspected before it is merged. \texttt{merge} performs the algebra
of \S\ref{sec:operator} and commits, recording both parent share identifiers on
the transaction.

Three implementation choices are worth naming because they affect what the
evaluation can claim.

\paragraph{The shapes come from the bundle, not the store.} Cardinality is read
from the shapes carried inside the share being merged, not from the local
store's copy. A peer's constraints therefore travel with its data, and merging
does not silently reinterpret their graph under our schema.

\paragraph{Conflict is reported at slot granularity, with its evidence.} A
\texttt{DecisionRecord} carries the subject, the predicate, the declared
\texttt{maxCount}, and the base, ours and theirs value sets. The person
resolving it is not told that a conflict exists and left to find it; they are
told which constraint made it one and what the three candidate states were.

\paragraph{The evaluated operator is the shipped one.} The benchmark of
\S\ref{sec:synthetic} contains its own reference implementation of every arm,
including ours, so that all eight are scored under one accounting contract. The
replay of \S\ref{sec:replay} deliberately does \emph{not}: it drives
\texttt{share\_merge::\{status,merge\}} through real bundles built by the
production share path. This matters because a benchmark's copy of an operator
can agree with a benchmark's copy of an oracle for reasons that have nothing to
do with the deployed system.

\section{Arm A: a synthetic divergence benchmark}
\label{sec:synthetic}

\paragraph{Design.} One seeded base graph over a synthetic vocabulary is copied
twice and edited independently --- separate random streams, so neither side's
edits can be a function of the other's --- under a six-operation model: assert,
retract, re-describe, re-type, alias-mint, node-add. An overlap parameter
controls how often an edit targets a contended quarter of the entity set, so
collision probability is set directly rather than left to chance. Eight
strategies then merge $(\base, \ours, \theirs)$ and are scored against one
oracle. There is no language model anywhere in the loop; the writers are
deterministic, so a run is its own ground truth.

\paragraph{The oracle is recomputable by a reader, with one stated exception.}
For the three triple-visible classes of \S\ref{sec:operator} (functional
divergence, cardinality overflow and delete/modify), \texttt{ground\_truth} derives the oracle
from the three graphs and the shapes and nothing else --- no edit log, no
generator intent --- so a reader holding only the published inputs recomputes
them exactly, and a generator bug cannot hide inside them. An instrument control
asserts precisely that and reports the count it recovered. The alias-mint class
is the exception, and necessarily so: two names for one entity is by
construction invisible in the triples (\S\ref{sec:limits}), so no oracle could
recover it from them. Those slots come from the generator's mint log. They are
$21$ of the $54$ conflicts summed over the five seeds, and \emph{every} arm in
Table~\ref{tab:armA} detects $0$ of them, ours included --- so the class bounds
the recall column and touches nothing else: no arm's precision or
true-positive count depends on a number the reader cannot recompute.

\paragraph{The accounting contract.} Every arm answers two questions whose
answers are in tension. \emph{Human decisions} is the number of slots it refused
to settle; a refused slot is held at its base value in the output, so no arm can
flag a conflict \emph{and} land a value for it. \emph{SHACL violations} counts
violations in what it landed automatically, against the same shapes that defined
the conflicts --- corruption it admitted without charging anyone. Either can be
driven to zero by sacrificing the other, so neither is reportable alone.
\emph{Triples lost} and \emph{triples spurious} are the third leg, and they
catch the arm that admits no corruption and asks no questions because it
discarded one side's work. Both are measured against the oracle's merge, in
which a true conflict is held at its base value. For an arm that asks, a change
it deferred on a slot that was not a true conflict is therefore counted as lost
or spurious: the change is waiting for a person, not destroyed.

\paragraph{Results.} Table~\ref{tab:armA} aggregates five seeds
($1, 7, 42, 99, 2026$) at $200$ entities, $200$ edits per side ($400$ in total)
and overlap $0.5$ --- the harness defaults, so the command below reproduces it
unchanged. Regenerate with \texttt{just bench merge -{}-seed <N>}; the outputs of those five
runs are committed under \texttt{benchmark/mergebench/results/}, and
Tables~\ref{tab:armAseed} and~\ref{tab:armAclass} are read from them. Instrument controls
with \texttt{just bench merge -{}-selftest} ($7$ properties, non-zero exit on any
failure).

\begin{table}[t]
\centering\small
\begin{tabular}{lrrrrrr}
\toprule
arm & decisions & precision & recall & SHACL & lost & spurious \\
\midrule
\texttt{git-turtle-reserialized} & 681--794 & 0.006--0.011 & 0.36--0.80 & 0--1 & 266--299 & 62--71 \\
\texttt{git-turtle-stable}       & 41--59   & 0.049--0.119 & 0.20--0.70 & 0--3 & 39--57   & 12--24 \\
\texttt{git-canonical}           & 106--144 & 0.028--0.057 & 0.30--0.70 & 0--3 & 95--128  & 18--31 \\
\texttt{union}                   & 0        & ---          & 0.00       & 53--72 & 2--7   & 112--136 \\
\texttt{lww-theirs}              & 0        & ---          & 0.00       & 0--1 & 7--16    & 4--10 \\
\texttt{triple-3way}             & 0        & ---          & 0.00       & 3--6 & 2--7     & 7--17 \\
\ctx                             & 159--182 & 0.025--0.046 & 0.50--0.80 & 0    & 128--150 & 63--72 \\
\op                              & 4--8     & \textbf{1.000} & 0.50--0.80 & \textbf{0} & \textbf{0} & \textbf{0} \\
\bottomrule
\end{tabular}
\caption{Eight strategies, five seeds, ranges across seeds. Recall is bounded
above by $1$ minus the alias-mint fraction for \emph{every} arm
(\S\ref{sec:limits}). The line-merge arms shell out to real \texttt{git
merge-file}; where git is absent they report \emph{unavailable}, never zero.
\ctx is our reimplementation of the Context Merge overlap rule, counted per
slot, not an execution of Quit Store. For arms that ask, \emph{lost} and
\emph{spurious} include changes deferred to a person on slots that were not
true conflicts; they are not irreversible deletions.}
\label{tab:armA}
\end{table}

\begin{table}[t]
\centering\small
\begin{tabular}{rrrrrr}
\toprule
seed & oracle & alias-mint & triple-visible & \ctx decisions & \op decisions \\
\midrule
1    & 14 & 7  & 7  & 159 & 7 \\
7    & 8  & 4  & 4  & 163 & 4 \\
42   & 12 & 4  & 8  & 173 & 8 \\
99   & 10 & 4  & 6  & 168 & 6 \\
2026 & 10 & 2  & 8  & 182 & 8 \\
\midrule
total & 54 & 21 & 33 & 845 & 33 \\
\bottomrule
\end{tabular}
\caption{Arm A per seed. Both arms detect every triple-visible conflict and no
alias-mint conflict at every seed.}
\label{tab:armAseed}
\end{table}

\begin{table}[t]
\centering\small
\setlength{\tabcolsep}{4pt}
\begin{tabular}{lrrrr}
\toprule
arm & alias-mint & card.\ overflow & delete/modify & functional div. \\
\midrule
\texttt{git-turtle-reserialized} & 0/21 & 8/10 & 9/10 & 13/13 \\
\texttt{git-turtle-stable}       & 0/21 & 0/10 & 10/10 & 13/13 \\
\texttt{git-canonical}           & 0/21 & 1/10 & 10/10 & 13/13 \\
\texttt{union}, \texttt{lww-theirs}, \texttt{triple-3way} & 0/21 & 0/10 & 0/10 & 0/13 \\
\ctx                             & 0/21 & 10/10 & 10/10 & 13/13 \\
\op                              & 0/21 & 10/10 & 10/10 & 13/13 \\
\bottomrule
\end{tabular}
\caption{Conflicts detected of conflicts declared, by class, summed over the
five seeds. The three right-hand columns are definitional for \op: its
predicate and the oracle read the same shapes. Of its $33$ decisions, $13$ are
functional divergence, $10$ are cardinality overflow on a predicate bounded at
four, and $10$ are delete/modify.}
\label{tab:armAclass}
\end{table}

Three readings matter.

\paragraph{The comparison with the nearest neighbor is not about detection.}
Summed over the five seeds the oracle contains $54$ conflicts, of which $21$ are
alias-mint and $33$ are visible at triple level. \op raised exactly those $33$,
with zero false positives at every individual seed. \ctx raised the
\emph{same} $33$ --- identical recall, seed by seed --- and $812$ others, for
$845$ questions in total. Two qualifications belong beside that figure. The
\ctx arm is our reimplementation of the overlap rule, and it counts one question
per changed slot on a subject both sides touched; Quit Store presents its
overlap as a review set rather than a question count, and a reviewer who settles
a whole subject at once would face fewer prompts than $845$. And the ratio
depends on how much the two sides overlap: rerunning the same five
seeds at overlap $0.1$, $0.3$, $0.7$ and $0.9$ gives $821$ questions against
$31$, $797$ against $25$, $899$ against $60$ and $983$ against $89$, a ratio
between $11\times$ and $32\times$. In those twenty further runs \op raised all
$205$ triple-visible conflicts with no false positive, while \ctx missed $5$ of
them; the identical detection reported above holds at overlap $0.5$ and is not a
general property of the overlap rule
(\texttt{benchmark/mergebench/results/overlap-*/}). The schema does not find conflicts the node-overlap
heuristic misses. It distinguishes the ones that need a person from the $812$
that do not, which is a $25.6\times$ difference in the quantity the operator is
actually spending: attention.

\paragraph{The zero columns are not all worth the same.} \texttt{union},
\texttt{lww-theirs} and \texttt{triple-3way} each report $0$ human decisions, and
none of them resolved anything. \texttt{union} admitted $53$--$72$ SHACL
violations. \texttt{lww-theirs} admitted almost none --- and silently dropped
$7$--$16$ triples and fabricated $4$--$10$, which is what last-writer-wins looks
like when you count what it discarded rather than what it complained about. Only
\op holds all three at zero while still asking.

\paragraph{A hypothesis of ours did not survive.} We predicted that
canonicalization alone would materially beat non-canonical line merge. Against
the re-serialized arm it does: unparseable output lines go from $116$--$169$ to
$0$. Against a \emph{stably} serialized Turtle file it does not.
\texttt{git-canonical} raises \emph{more} conflicts ($106$--$144$) than
\texttt{git-turtle-stable} ($41$--$59$), because sorting scatters each subject's
triples into one sorted run where two sides' additions land adjacent and collide,
whereas subject-grouped Turtle absorbed them into a block. The honest statement
is narrower than the hypothesis: canonicalization buys syntactic integrity and
independence from serialization order, and it does not by itself reduce conflict
count. It is a trade, and we report it as one.

\paragraph{The circularity, stated rather than managed away.} \op and the oracle
both read the same shapes graph, and for the triple-visible classes their rules
coincide by construction. \op's precision of $1.000$ on this arm is therefore a
property of the design, not evidence about it. We do not repair this by making
the oracle cleverer, which would only move the coincidence somewhere less
visible. We handle it in two ways: by reporting per-class results
(Table~\ref{tab:armAclass}) so a reader can see which rows are definitional, and
by confining this arm's contribution to the \emph{cost of the alternatives}. The replay arm (\S\ref{sec:replay}) replaces the generated
corpus with recorded divergence and the harness with the shipped operator, but
its positive results are scored against the same shapes. The evidence that does
not pass through the shapes is a separate, smaller arm (\S\ref{sec:oracle}).

\paragraph{The shipped planner differs from the reference arm on one slot.}
\op in this benchmark is a reference implementation. The harness can also call
the library's own slot planner (\texttt{unified-core}, built with
\texttt{-{}-features shacl,merge-benchmark}). On four seeds the two are
identical on every metric. On seed $1$ they differ on one slot, bounded at four
values, where one side is unchanged and the other alone holds five. The
reference arm flags a conflict, as the design rule requires. The planner
classifies the change as one-sided and offers no decision, so in the benchmark,
which calls the planner without the store's write gate, it scores $6$ true
positives where the reference scores $7$ and admits one violation. Through the
shipped paths the five values are never written: a store with the shape loaded
refuses the write that would create that side, and when we built such a bundle
by tightening the bound after the write, \texttt{share\_merge} refused the
whole merge and wrote nothing. The defect is therefore a missing decision, an
all-or-nothing refusal where the rule calls for a question, and not a silent
violation. The figures in Tables~\ref{tab:armA} and~\ref{tab:armAclass} are the
reference arm's.

\paragraph{Wrong shapes: what the circularity hides.} A second judge derives the truth from the three graphs
and a separate declaration of each predicate's intended cardinality, and never
reads the shapes. With the shapes as published, its truth equals the
generator's on every seed and every count in Table~\ref{tab:armA} reproduces.
That agreement is by construction, because the shapes state the intent, and we
do not count it as new evidence. The measurement is what happens when the
operator is told shapes that are wrong while the truth stays fixed
(Table~\ref{tab:wrongshapes}). A missing bound costs recall and admits
violations of the intended cardinality into the merged graph, while precision
stays at $1.000$: the operator looks perfect on what it flags and is silent on
what it was never told. A bound that is too tight or should not exist costs
precision, as extra questions, and corrupts nothing. The shipped planner
raises far fewer under such bounds than the reference arm, $14$ false
positives against $278$. We do not credit it for that: it considers a slot only
when both sides changed it, which is the same classification that misses the
one-sided overflow above. \ctx reads no shapes and is unchanged in every row, at $812$
false positives. Shape quality therefore transfers directly into merge quality,
and asymmetrically.

\begin{table}[t]
\centering
\footnotesize
\setlength{\tabcolsep}{3pt}
\begin{tabular}{lrrrrrr}
\toprule
 & \multicolumn{3}{c}{\op} & \multicolumn{3}{c}{\texttt{unified-core}} \\
operator's shapes & P & R & viol. & P & R & viol. \\
\midrule
as published & 1.000 & 0.611 & 0  & 1.000 & 0.593 & 1 \\
\texttt{tag} unbounded & 1.000 & 0.426 & 10 & 1.000 & 0.426 & 10 \\
\texttt{label}, \texttt{status} unbounded & 1.000 & 0.389 & 8 & 1.000 & 0.370 & 9 \\
\texttt{tag} at most 2 & 0.106 & 0.611 & 0  & 0.696 & 0.593 & 1 \\
\texttt{note} at most 1 & 0.194 & 0.611 & 0  & 0.842 & 0.593 & 1 \\
\bottomrule
\end{tabular}
\caption{Arm A re-judged without reading the shapes, pooled over five seeds
($54$ conflicts, $21$ of them alias-mint, which caps recall at $0.611$).
\emph{viol.} counts intended-cardinality violations left in the merged graph. The judge is
\texttt{benchmark/mergebench/judge.py}.}
\label{tab:wrongshapes}
\end{table}

\section{Arms B and C: recorded divergence, shipped operator}
\label{sec:replay}

The synthetic arm can measure what the alternatives cost. It cannot be evidence
that our operator is correct, because its oracle and its operator share a
definition. This arm removes one half of that circularity and not the other. The
incidents are \emph{recorded} production divergence rather than generated: which
names were later declared one entity, and which subjects carried doubled values
and how many. The operator under test is the shipped \texttt{share\_merge} driven
through real bundles, so a disagreement between operator and expectation is a
finding rather than a harness bug. The triples that carry each incident into a
merge are constructed by the replay harness from those identities and
multiplicities, with placeholder values; this arm replays recorded incidents, not
recorded edit histories. The judgement, however, is still derived from the shapes. A
\texttt{comment-double} decision counts as correct only because the description
predicate carries \texttt{sh:maxCount 1}, and \emph{SHACL violations} is scored
against the same shapes in every arm. The one result in this arm that does not
depend on the shapes is the alias column, and it is a negative: $0$ of $93$.
An oracle that does not read the shapes judges the same operator on a small
set of cases in \S\ref{sec:oracle}.

\paragraph{The corpus.} Extracted from a live multi-agent knowledge graph in
which software agents and people had been writing concurrently for months, and
committed as a fixture. Entity names are replaced by random labels and the
mapping is discarded at build time; structure, cardinality and class membership
are preserved exactly, and injectivity is verified over the $1{,}076$ names.
Two divergence classes were recorded.

\emph{Alias-mint}: two names silently minted for one entity, later repaired by a
person asserting \texttt{owl:sameAs}. There are $171$ raw \texttt{sameAs} edges,
which reduce to $105$ distinct undirected repairs ($66$ edges were one repair
recorded twice). Each is one repair a person actually performed --- the
manual-effort baseline. \textbf{The class splits, and the split matters}: $52$
of the $105$ are \emph{id-form}, two spellings of the same commit identifier,
which a normalization rule could retire without asking anyone; the other $53$
are two English phrasings of one concept, where nothing but a person can decide.
Reporting $105$ as one undifferentiated ``human decisions'' figure would
overstate the irreducible cost roughly twofold.

\emph{Comment-double}: a \texttt{sh:maxCount 1} description predicate that the
production append path doubled instead of replacing --- $939$ affected subjects
carrying $1{,}784$ excess values.

\paragraph{Results.} Table~\ref{tab:armB} reports the replay. Regenerate with
\texttt{cargo run -{}-example replay -{}-features shacl}.

\begin{table}[t]
\centering\small
\begin{tabular}{lrrl}
\toprule
scenario & decisions & historical & outcome \\
\midrule
\texttt{alias-id-form}   & 0   & 46 & merged --- \textbf{false negative} \\
\texttt{alias-semantic}  & 0   & 47 & merged --- \textbf{false negative} \\
\texttt{comment-double}  & 939 & 0  & conflicts --- raised as decisions \\
\texttt{sameas-repair}   & 0   & 20 & merged --- repair survives (wanted) \\
\bottomrule
\end{tabular}
\caption{Recorded divergence replayed against the shipped operator. $12$ pairs
($6$ id-form, $6$ semantic) were excluded as \emph{chained} --- sharing an
endpoint with a pair already kept --- because including them measures an incidental
collision on the shared endpoint rather than alias detection.}
\label{tab:armB}
\end{table}

\paragraph{The operator is blind to the entire alias class: $0$ of $93$.} This
reproduces the synthetic arm's $0$ of $21$ on alias pairs a person actually
repaired, and it is the honest false-negative column. The zero follows from the
structure of the case, since two different names share no slot, so it
demonstrates a limit rather than measures a rate. Two names for one entity is
invisible to \emph{any} triple-level operator, shape-aware included; no schema
fixes it, because it is a limit of the medium the operator works in.

\paragraph{The operator raises $939$ decisions on a class production surfaced
none of.} The same class produced $1{,}784$ silent excess values through the
append path. This is a counterfactual and must be read as one
(\S\ref{sec:limits}).

\paragraph{A repair survives a concurrent edit.} \texttt{owl:sameAs} carries no
cardinality constraint, so a repair asserted on one side unions rather than
racing an edit to the aliased node on the other. A merge that could drop repairs
would make the repair path itself unreliable, so this outcome is the wanted one
rather than an incidental one.

\paragraph{The negative control.} Rerunning with the description predicate's
\texttt{maxCount} unbound takes \texttt{comment-double} from $939$ decisions to
$0$, everything merging cleanly. The decisions are therefore \emph{shape-derived}
--- not manufactured by the harness --- and this control arm doubles as the
closest thing to a simulation of the production append path: it reproduces the
doubling. That is the paper's thesis stated as an experiment. Unbind the
constraint and the conflict ceases to exist; the schema is what made it one.

\paragraph{Arm C: share, diverge, reconnect.} A case study walks the full
lifecycle on a real bundle and verifies all eight hops: share a base bundle;
a peer publishes a divergent share naming it as parent; we edit locally; status
before reconnect reports \texttt{diverged} with $+40$ triples each side and $0$
conflicts; the merge asserts $40$ and retracts $0$; the transaction records two
parents; both sides' work is present afterwards ($40/40$ each); and status after
reconnect reports nothing outstanding. One detail is worth carrying: the graph
hash is stable across runs while the share identifier is not, because the
manifest embeds a timestamp. Fork-point anchoring depends on the graph hash, so
lineage is unaffected --- but a share identifier must not be used as a content
identity.

\subsection{Arm D: pairwise identity proposals}
\label{sec:armD}
A preregistered follow-up tests whether Jev~1.13.0, TypeSafe AI's hosted
decision model that returns typed answers and probabilities rather than
generated text~\cite{typesafe2026jev}, can propose identity decisions for the
$93$ aliases missed by the triple-level operator. Given a pair, it
returns \texttt{same}, \texttt{different}, or \texttt{cannot\_tell}; the latter
and confidence below $.75$ abstain. A proposal never applies an identity merge.
The free comparators use normalized label edit similarity, MiniLM cosine, their
union (all at $.90$), and a conservative explicit repository/commit-ID control.
Thresholds and prompts were frozen before paid inference.

Evidence comes from preserved ROOT assertions before each pair's earliest
recorded transitive identity connection, with both transaction and valid-time
restrictions. Identity annotations are stripped; current text and vectors do
not fill missing historical evidence. Only $60/93$ positives have mechanically
eligible evidence ($39/46$ id-form, $21/47$ semantic). All original positives
remain in recall; we also report recall among eligible positives.

Same-type neighbors ranked by maximum label/cosine score supplied proposed
negatives, independently adjudicated using distinctness evidence. There are
$114/46=2.478$ negatives per id-form positive and $41/47=0.872$ per semantic
positive, short of the respective $3{:}1$ targets by $24$ and $100$.
Overall $155$ negatives exceed the $93$ floor but miss the $279$ target by $124$;
one semantic negative exceeds the request-size cap. The reviewed semantic pilot
negatives have clearly different descriptions, limiting hard-negative claims.

\paragraph{Pilot and availability.}
The ten-item pilot produced eight typed responses and two client timeouts.
After a first timeout at $30.158$ seconds, a registered amendment raised the
client deadline from $30$ to $120$ seconds. Four responses took
$0.328$--$1.418$ seconds, followed by another timeout at $120.115$ seconds.
The first latency uses attempt time to durable error-receipt time; subsequent
measurements use a monotonic timer. Size alone does not explain these events:
$4{,}355$ bytes timed out, $5{,}427$ bytes succeeded in $0.344$ seconds, and
$5{,}538$ bytes timed out. A second public amendment allowed two retries for
only the four untouched pilot negatives; all four responded on their first
attempt. Earlier failures were retained, not retried. First-attempt and eventual
item availability are both $8/10$. The pilot recovered $2/3$ id-form and $1/2$
semantic positives, with one unavailable positive and one abstention; its four
responding negatives produced no false identity proposals.

The held-out run used $204$ untouched eligible items, at most three attempts per
item, $120$ seconds per attempt, a $612$-attempt cap and a \$0.20 allowance.
The first-$50$ stop gate passed: first-attempt availability $48/50$, eventual
availability $50/50$, requiring $52$ attempts. Valid abstentions count as
available responses but not recovered aliases.

The completed held-out run used $216$ attempts; first-attempt
availability was $192/204$ and eventual item
availability $204/204$. Including the pilot, returned
usage implies a list-price subtotal of \$0.014897148;
14 attempts have unknown possible charges. Total account spend
is therefore unknown. The local guard reserved \$0.001 per unknown or in-flight
attempt in addition to known usage; this is not a certified debit bound.

\paragraph{Held-out results.}
Table~\ref{tab:armD} keeps pilot items separate and retains unavailable evidence
in the original-positive denominator. False proposals are the errors that would
collapse distinct entities if accepted; actual merges are zero in every arm.
\begin{table}[ht]
\centering\small
\begin{tabular}{llrrr}
\hline
Stratum & Method & Recall, original & Recall, eligible & False proposals\\
\hline
id-form & Jev & $10/43$ & $10/36$ & $0$\\
id-form & Label & $0/43$ & $0/36$ & $0$\\
id-form & Cosine & $0/43$ & $0/36$ & $2$\\
id-form & Union & $0/43$ & $0/36$ & $2$\\
id-form & Commit-ID & $10/43$ & $10/36$ & $0$\\
semantic & Jev & $1/45$ & $1/19$ & $1$\\
semantic & Label & $3/45$ & $3/19$ & $0$\\
semantic & Cosine & $0/45$ & $0/19$ & $0$\\
semantic & Union & $3/45$ & $3/19$ & $0$\\
semantic & Commit-ID & $0/45$ & $0/19$ & $0$\\
\hline
\end{tabular}
\caption{Arm D held-out pairwise decisions at the registered thresholds.
Original-positive recall includes missing evidence; eligible recall excludes
those preparation failures but retains model abstentions and unavailable calls.}
\label{tab:armD}
\end{table}

Across pilot and held-out items, Jev recovered $14/93$ aliases; this is
partial recovery of the operator's $0/93$ gap, not end-to-end entity resolution.
On held-out positives it recovered $11/88$ overall, with
one false identity proposal. The gains are predominantly id-form: the
commit-ID control recovers ten held-out aliases without paid inference. On
semantic positives Jev recovers one, versus three for label matching; the
pilot is too small to reverse that conclusion. The preregistered requirement to beat the free baselines fails on the semantic
stratum: this is a negative finding for Jev at the registered operating point,
not evidence for adding a paid resolution step.
Precision, abstention rates and threshold curves are released in the offline
artifact; shared endpoints and selected negatives limit population inference.

Held-out Jev precision is $10/10$ for id-form and $1/2$ for semantic proposals.
Among eligible items, decision abstentions are $43/147$ id-form and $21/57$
semantic; another $7$ and $27$ items respectively were withheld during evidence
preparation. The ten ID-form recoveries overlap the commit-ID control on only
five pairs: equal recall does not mean identical decisions.

The false proposal illustrates the distinction the prompt asks for: one record
names CPU thermal excursions, while the other names missing fan telemetry on
the same host. They describe different failure modes even though their host and
thermal context overlap. Jev proposed identity at confidence $.77$; both free
similarity scores were below $.90$. This observed error, together with lower
semantic recall than label matching, argues for reviewed proposals rather than
automatic identity collapse. Structured-predicate checks found zero identity
predicates in all $248$ frozen request states, with an injected-edge positive
control detected; this control does not certify arbitrary prose.

\section{Divergent shape declarations (Arm E)}
\label{sec:schema}

All earlier merge arms hold the shape contract fixed. Arm E instead tests a
research orchestration: merge the shape declarations as RDF under a fixed
SHACL-SHACL meta-shape, validate that contract, then merge data under it and
validate again. The actual library share-merge APIs perform both merges using
real common-base envelopes. The meta-shape declares path, bound and datatype
functional on property-shape nodes. Shape graphs have stable named nodes;
neither blank-node correspondence nor general SHACL meta-validation is tested.

Five explicit fixtures cover tightening \texttt{maxCount} from two to one,
loosening a one-value bound, adding a functional property, changing datatype
from string to integer and removing a shape. Each is repeated with five seed
identities. Left edits under the old contract and right under its changed
contract; both source branches are checked valid under their own contracts.
An independent positive control changes the same bound differently in both
shape branches and requires a meta-conflict. Positive and negative validation
controls prove the meta-contract is exercised.

\begin{table}[ht]\centering\small
\begin{tabular}{lrrrrrrrr}\toprule
Case & TP & FP & FN & Merge & Context & Lost & Extra & SHACL \\ \midrule
tighten & 1 & 0 & 0 & 1 & 1 & 0 & 0 & 0 \\
loosen & 0 & 0 & 0 & 0 & 1 & 0 & 0 & 0 \\
property-added & 1 & 0 & 0 & 1 & 1 & 0 & 0 & 0 \\
datatype & 0 & 0 & 1 & 0 & 1 & 0 & 2 & 1 \\
removed & 0 & 0 & 0 & 0 & 1 & 0 & 0 & 0 \\
\bottomrule\end{tabular}
\caption{Arm E, per seed. The datatype incompatibility is an oracle conflict missed by the cardinality detector and caught by post-merge SHACL validation. These are normalized automatic-output metrics, not partial deployment results.}
\label{tab:schema}\end{table}

The cardinality cases behave as the merged contract specifies, including
conflicts appearing on tightening and dissolving on loosening/removal. Datatype
evolution exposes a boundary: the shipped detector reads cardinality bounds,
so it misses the incompatible left-side string in the new integer contract.
Post-merge validation detects the violation. That rejection is not counted as
conflict recall, and the invalid candidate must not be admitted by a pipeline
that treats validation as a gate. Merely merging shape triples does not derive
a complete conflict policy for every SHACL constraint.

Table~\ref{tab:schema} uses explicit fixture oracles and holds a conflicted slot
at base for automatic-output accounting, matching Arm A. The library itself
blocks the entire merge; this normalization is not partial application. The
Context column is the same changed-subject overlap rule on these literal-only
fixtures, not an execution of Quit Store. Five identities do not constitute
five sampled schema-evolution workloads. The orchestration is an experimental
driver, not a deployed schema synchronization or negotiation feature.
Regenerate with \texttt{just bench schema} and \texttt{just bench schema-report};
\texttt{benchmark/shapes/} publishes all three input graphs, changed contracts,
expected outputs and actual results.

\section{Arm F: an oracle that does not read the shapes}
\label{sec:oracle}

Every positive result so far is scored against the shapes that also define the
conflict. This arm removes that dependence for a small set of cases. Each case
is a fixture that states, for one slot, the base, ours and theirs value sets,
the slot's intended cardinality, the decision a person takes, and the expected
outcome: the current facts, the history records with their valid intervals, and
the ordered transaction trace. The judge compares those expectations with what
the store holds after the merge. It reads no shapes graph and imports no merge
code; the expectations were written from the design document before the adapter
existed. A driver builds each run from the case inputs alone and never reads
the expected outcome.

There are $24$ cases: clean one-sided and two-sided changes, convergent
changes, deletions, each decision on a contended slot (ours, theirs, base, a new
value, a refused attempt to keep both, and a deferred decision), and three
transaction-time and valid-time reads of a decided merge. We ran them against
two shipped paths in private scratch stores: \texttt{share\_merge}, the operator
of \S\ref{sec:replay}, and the fork-promote path that shares its slot planner.

\begin{table}[t]
\centering
\small
\begin{tabular}{lrr}
\toprule
 & share merge & fork promote \\
\midrule
facts, history and trace agree            & 19 & 22 \\
facts and history agree, trace differs    &  4 &  0 \\
facts or history disagree                 &  0 &  0 \\
case not reachable through the path       &  1 &  2 \\
\midrule
conflict report agrees (of cases with one) & 10/10 & 10/10 \\
\bottomrule
\end{tabular}
\caption{Arm F: $24$ single-slot cases judged by an oracle that reads no
shapes. Regenerate with \texttt{benchmark/usecase-oracle/run\_native.py}; its
README gives the two invocations.}
\label{tab:oracle}
\end{table}

On every case either path can reach, the facts and the history agree with the
oracle (Table~\ref{tab:oracle}). The four trace differences on the share path
have one cause. When a merge has nothing to apply, the operator still records
one empty receipt transaction, and the fixtures expect no transaction. The
design document states both, in different places; we report the disagreement
as measured and did not change the fixtures to match the code. The case
neither path reaches needs an empty base, which the adapter cannot write. The
second unreachable case on fork-promote is a real difference between the paths:
when both sides converge on a value, fork-promote records no conflict, so a
person cannot then choose a different value, and the share path can.

As a control, four captures were scored against a different case's
expectations, and all four failed, so agreement is not an artifact of a judge
that accepts anything.

\paragraph{What this arm does and does not establish.} It is the first evidence
in this paper that does not pass through the shapes: the shipped operator
produces the facts and history that an independently stated expectation asks
for. It is small. The cases are single-slot and synthetic, written by the
project that wrote the operator. List-valued slots and repeated merges are
specified as fixtures but have no adapter yet, and the conflict report is
checked only on contended slots. No case has one side alone exceed a bound, so
this arm does not exercise the planner difference of \S\ref{sec:synthetic}. It does not re-judge the $33$ decisions of
\S\ref{sec:synthetic}.

\section{Limitations}
\label{sec:limits}

\paragraph{Scope: less human attention, not general merge correctness.} The
claim this paper supports is narrow. For slot-level cardinality conflicts, the
operator asks a person fewer questions than the alternatives we measured, and it
loses nothing silently. It is evaluated in one deployment, over one ontology
family, with SHACL cardinality bounds as the only constraint that defines a
conflict. It does not cover alias divergence ($0$ of $93$ detected), datatype or
other SHACL constraints, lists and blank-node structures, or ontologies that
diverge beyond the Arm E fixtures. Every precision and true-positive figure in
\S\ref{sec:synthetic} and \S\ref{sec:replay} is relative to the shapes that also
define the conflicts, so none of them is evidence of shape-independent
correctness. The only such evidence is Arm F (\S\ref{sec:oracle}): $24$
single-slot synthetic cases, written by the same project, with list-valued
slots and repeated merges not yet covered. It does not re-judge the benchmark
or the replay.

\paragraph{Alias divergence is invisible, and that is a property of the medium.}
When two writers mint different names for one entity, the graphs disagree about
the world while agreeing about every triple. No triple-level operator can see
it, so a recall of $1.000$ is not achievable by anything in Table~\ref{tab:armA}.
We left the class in the benchmark deliberately: it is a real
divergence defect in the recorded corpus ($105$ repairs), and excluding it would have produced a
shape-aware arm scoring $1.000$ on both precision and recall as an artifact of
what we chose to measure. Arm D tests a separate pairwise entity-resolution stage, using labels and
historical neighborhoods to propose identity decisions. Its held-out run recovered
$11/88$ aliases ($11/55$ with eligible evidence) and proposed one false identity.
Semantic recall was $1/45$ ($1/19$ eligible), compared with label matching's
$3/45$ ($3/19$ eligible) and zero false proposals. The preregistered baseline-beating
condition fails on this stratum, a negative finding for Jev. Candidate generation
is a separate requirement whose recall bounds end-to-end resolution, and we
measured it with free signals only. Each pair's left endpoint ranks every entity
asserting one of its types before the repair, by the larger of normalized label
similarity and MiniLM cosine over the same cleaned pre-repair evidence, with ties
counted against the true partner. Recall at $k = 1, 5, 10, 50$ is $38$, $54$,
$56$ and $57$ of $93$ (id-form $23, 32, 34, 35$ of $46$; semantic
$15, 22, 22, 22$ of $47$). The ceiling is the type block, not the ranking:
$30$ partners cannot be ranked at all, $26$ of them because they never enter the
anchor's typed pool ($15$ had no pre-repair facts and $11$ carried disjoint
types), and $25$ of the $30$ are semantic. Among the $63$ rankable partners,
$56$ fall in the top ten, and cosine alone puts more partners first ($43$) than
the combined score ($38$). A resolver fed by this blocker would lose about a
third of all aliases, and over half of the semantic class, before pairwise
resolution begins. Historical evidence gaps, the semantic negative
shortfall, easy reviewed negatives, and shared endpoints limit generalization.
No actual identities were merged, so the study measures proposed errors rather
than downstream damage or the economic cost of accepting them.

\paragraph{The $939$ is a counterfactual.} Those doublings arose from sequential
re-posts through a single write path, not from a share, a divergence and a merge.
The supported claim is narrow: had the same edits arrived as a divergence and
been reconciled by this operator, each would have been surfaced as a decision
instead of silently doubling. It is \emph{not} a claim that the operator was
running and prevented $1{,}784$ doublings.

\paragraph{Comment bodies in the merge replay corpus are synthetic.} Real values are prose
about a deployment. In arm B, only the incident identities and the multiplicity are real --- which is all the merge
reasons about, but it does mean the corpus cannot support any claim about value
content.

\paragraph{A conflict blocks the whole merge, not the conflicting slot.} The
operator returns \texttt{conflicts} with nothing asserted. On a slice like the
$939$-subject one, a single reconnect writes nothing until every contended slot
is resolved. This is the conservative choice and it is a real usability cost;
partial application with a held-back conflict set is future work.

\paragraph{Ground graphs only.} The benchmark's graphs contain no blank nodes,
so canonicalization there reduces to sorting. RDFC-1.0 is the substrate we build
on and is not exercised by this arm; nothing here is evidence about blank-node
canonicalization.

\paragraph{Timing is indicative.} Arm A uses one machine, one process and single runs. The
line-merge arms include process spawn for \texttt{git merge-file}, of order
$5$\,ms, which dominates their timing at every size measured; those columns are
not comparable with the in-process arms and we do not report them as if they
were.

\paragraph{Slot granularity is an assumption.} A conflict is reported over a
$(subject, predicate)$ slot because that is the unit \texttt{sh:maxCount} is
declared over. An operator working at another granularity would need a different
scorer, and the comparison in Table~\ref{tab:armA} would not transfer unchanged.

\paragraph{Evolving shapes need more than cardinality.} Arm E uses a small,
fixed meta-contract over named shape nodes and an experimental orchestration,
not a production schema negotiation protocol. Its datatype case is a detector
false negative detected by validation; passing the shape-graph merge does not
make all data changes under the new contract safe.

\paragraph{One deployment.} The recorded corpus comes from a single production
store. Its class distribution --- alias-mint and functional-value doubling
dominating --- may not generalize, and we make no claim about divergence rates
in other systems from it.

\paragraph{How often production writes would have needed a merge is only sampled.}
The deployment writes through one path, so its history contains overwrites and
no merges. We counted the overwrites that look like concurrent edits in a
read-only copy of the store: a single-valued slot replaced by a different value
within $300$ seconds by a different declared writer. The count is a sample, not
a census. It covers $299{,}026$ of $1{,}443{,}490$ transactions ($20.7\%$): the
first $99{,}048$ consecutively, then one $2{,}000$-transaction window from each
of $100$ equal strata. It found $10$ such pairs in the prefix and $1{,}344$ in
the sample, almost all on work-item status, version and modification-time
slots. In $1{,}250$ of the sampled pairs both writes carry the same recorded
principal behind different declared names, and $89$ have different principals.
These are physical value replacements; the base each writer read is not
recorded, so none is shown to be a lost update, and uncovered history is
unknown rather than zero (\texttt{scripts/merge-shadow-census.py}).

\paragraph{Private evidence limits inference reproducibility.}
Arm D publishes the item labels, frozen free-baseline scores, prompt, numeric
responses, attempt ledger, and offline scorer. Its structural item views replace
private strings with random tokens and withhold descriptions; selected evidence
examples are scrubbed editorial paraphrases. These artifacts reproduce scoring,
not the original model requests or embeddings. Reviewer inspection used the
preserved private evidence. A reader cannot independently regenerate the model
outputs from the public views, and a commercial model may change separately.
Timeouts may be billed without usage responses, so list-price subtotals do not
establish total account spend. First-attempt and eventual-item availability
must be reported separately when retries are allowed.

\section{Related work}
\label{sec:related}

\paragraph{Git-backed RDF versioning.} Quit Store~\cite{arndt2019quit} is the
closest neighbor and the reason this paper makes no priority claim. It places
RDF under git --- the system's own gloss on its name is ``Quads in Git'' ---
versions quads with commit-level provenance, and supplies merge strategies,
including a Context Merge that identifies conflicts by node overlap and hands
them to a person. The triple-set framing of three-way merge is present there; we
are not introducing it.

Our delta is where the conflict predicate comes from. Context Merge derives
contention from the \emph{shape of the change}: it marks the subject and object
of every added or removed statement with the commit it came from, and a node
marked by both sides becomes a conflict. We derive it from the \emph{shape of
the data}, as declared in SHACL. The difference is a deliberate design choice on
their part rather than an oversight, and they state it: Quit Store declines to
encode semantic rules and relies on the pure RDF data model instead, and it
presents Context Merge as a \emph{supervised} approach to identifying conflicts
rather than as a decision procedure. Our claim is correspondingly narrow. SHACL
is exactly the source of semantic rules they chose not to hard-code, and what
lets us consult it is circumstance rather than insight: in a shapes-governed
store those rules are already present, already authored by the people who
authored the data, and already the contract the graph is validated against.

They go one step further than declining, and it is worth stating plainly because
it is the closest any prior work comes to the idea in this paper: Quit Store's
discussion names SHACL itself, proposing that users define application-domain
rules as ontological restrictions and that adherence to them be enforced by
continuous integration over the repository. The \emph{placement} is then the
whole of our delta. There, shapes are a quality gate applied to a merge that has
already happened; here, shape cardinality is the conflict predicate that decides
what the merge is permitted to settle at all, and the post-merge check they
propose is additionally what we run to audit the result
(\S\ref{sec:operator}). This operator therefore does not introduce SHACL to RDF
merging. It moves it from after the merge to inside it, and the $845$-against-$33$
result below is the measurement of what that move buys.

The consequence is measurable rather than rhetorical, and it is the central
result of \S\ref{sec:synthetic}: across five seeds the two approaches detect an
identical set of $33$ triple-visible conflicts, and the context heuristic asks
$845$ questions where cardinality asks $33$. Overlap is a good proxy for ``these
edits are related'', which is what Context Merge sets out to find. It is a poor
proxy for ``these edits cannot both be true'', which is the smaller question that
actually needs a person --- and Context Merge does not claim to answer it. The
$845$ against $33$ therefore measures two different questions being asked, not
one strategy failing at the other's task: it quantifies what the extra
information buys, and that information is only available to a merge that has a
shapes graph to consult. We position this operator as an extension of that line
of work: the algebra is theirs, the conflict predicate is the contribution.

\paragraph{Diff and patch over versioned graphs.} TerminusDB~\cite{terminusdb} provides diff and
patch over a versioned graph store. RDF Patch and LD Patch standardize the
description of change sets. These solve the representation and transport of a
delta; they are complementary to, and do not decide, what makes two deltas
irreconcilable.

\paragraph{Canonicalization.} W3C RDF Dataset Canonicalization~\cite{rdfc2024} (RDFC-1.0,
formerly URDNA2015) gives deterministic labeling of blank nodes and hence
decidable graph equality. We build directly on it: it is what makes ``the same
triple'' well defined across two independently serialized bundles, and it is a
substrate rather than a competitor. R43ples~\cite{graube2016r43ples} and archiving systems such as
OSTRICH~\cite{taelman2018ostrich} address efficient storage and querying of revision histories, a
different problem from reconciling two live divergent copies.

\paragraph{Convergent replicated types.} CRDT constructions for RDF, such as
SU-Set~\cite{ibanez2012suset}, guarantee convergence without coordination. They and this operator make
opposite trades deliberately. A CRDT resolves every case automatically, which
requires that some rule always picks a winner; for a functional predicate whose
two values are both asserted in good faith, any such rule silently discards
somebody's knowledge. We would rather stop and ask, and accept that this means
a merge can fail. Where automatic convergence is required, a CRDT is the right
tool and this is not; where the graph is a shared record of what is believed to
be true, a silently dropped assertion is the more expensive failure. Operational
transformation has the same shape of trade in a different lineage.

\paragraph{Schema-aware merging elsewhere.} Ontology alignment and matching
address a related question --- reconciling different conceptualizations --- but
operate on schema-to-schema correspondence rather than on three-way
reconciliation of instance data under a shared schema. Structured merge tools
for source code (semi-structured and AST merges) share our motivation exactly:
raise the merge above lines to the level at which the medium defines conflict.
Our claim is the RDF instance of that principle, with SHACL supplying the
structure.

\paragraph{Evidence maintenance and reviewed agent memory.}
Choudhary's PolyGraph~\cite{choudhary2026receiptstrustworthycleanupagentbuilt}
records alternative evidence bundles for derived artifacts and audits receipts
by re-derivation after evidence removal. Its abstract reports that deleting only
directly extracted facts leaves substantial contamination. This is complementary
to shape-aware merge: structural consistency does not establish that an
artifact's evidence remains sufficient after source withdrawal. We do not
benchmark receipt fidelity here.
Raj's Plant Memory~\cite{raj2026plantmemorytechnicallanguage}
combines rule-first extraction, optional LLM review, reified facts with evidence,
confidence and verification state, technician review, and MCP access. Its abstract
reports a synthetic 400-work-order evaluation. It is an application-specific
reviewed memory system; our operator reconciles divergent RDF graphs under shared
SHACL constraints. Neither abstract establishes a comparison of temporal or
withdrawal semantics with this implementation, and we make no such claim.

\section{Conclusion}

Merging knowledge graphs by merging their text asks a line-oriented tool to
infer a semantic property, and it fails in both directions: it manufactures
conflicts from serialization order and misses the divergences that genuinely
cannot be reconciled automatically. Taking the canonical triple as the unit of
change removes the first failure entirely. Taking the conflict predicate from a
SHACL shapes graph addresses the second, and does so without adding a policy
that has to be maintained alongside the ontology --- because it \emph{is} the
ontology, and the same shapes then audit the result.

The evaluation supports a narrow claim, which is the one we make. Against the
nearest neighbor, schema-derived conflict semantics do not detect more: they
detect the same $33$ triple-visible conflicts across five seeds while asking
$33$ questions rather than $845$, with no triples lost, none fabricated and no
constraint violations admitted against the same shapes. That is less human
attention for cardinality conflicts, not general merge correctness; an oracle that reads no shapes agrees with
the shipped operator on the facts and history of every reachable case among $24$ single-slot ones. Replaying scenarios built from $939$ recorded duplicate-value incidents and
$93$ of $105$ recorded repairs ($12$ chained pairs excluded) against the shipped
operator reproduces that benefit on incidents nobody generated for the purpose --- and reproduces the blind spot too:
alias divergence is invisible to triple-level merge, $0$ of $93$, and no schema
will fix it.

The schema-evolution trials refine that claim.
Merging declarations first changes cardinality decisions as intended in the
tested fixtures, but datatype incompatibility still needs validation and a
resolution policy. Schema-first orchestration is therefore not a complete
conflict-resolution rule on its own.

Three directions follow. The use-case oracle covers single slots; extending it
to lists, repeated merges and the benchmark's own decisions would judge the
operator independently of the definition it implements at scale.
Partial application would let a merge land its
uncontended slots and hold back only the contended ones, removing the cost
noted in \S\ref{sec:limits}. And the alias class, which the recorded corpus shows is real
and which this operator cannot see, is where entity resolution would have
to meet merge --- with the open question being whether its errors are cheaper
than the divergence it repairs.

\bibliographystyle{plain}
\bibliography{references}

@article{arndt2019quit,
  author  = {Arndt, Natanael and Naumann, Patrick and Radtke, Norman and
             Martin, Michael and Marx, Edgard},
  title   = {Decentralized Collaborative Knowledge Management Using {Git}},
  journal = {Journal of Web Semantics},
  volume  = {54},
  pages   = {29--47},
  year    = {2019},
  doi     = {10.1016/j.websem.2018.08.002}
}

@misc{rdfc2024,
  author       = {Longley, Dave and Kellogg, Gregg and Yamamoto, Dan},
  title        = {{RDF} Dataset Canonicalization ({RDFC-1.0})},
  howpublished = {W3C Recommendation},
  year         = {2024},
  month        = may,
  url          = {https://www.w3.org/TR/rdf-canon/}
}

@inproceedings{taelman2018ostrich,
  author    = {Taelman, Ruben and Vander Sande, Miel and Verborgh, Ruben},
  title     = {{OSTRICH}: Versioned Random-Access Triple Store},
  booktitle = {Companion Proceedings of The Web Conference 2018},
  year      = {2018},
  doi       = {10.1145/3184558.3186960}
}

@inproceedings{graube2016r43ples,
  author    = {Graube, Markus and Hensel, Stephan and Urbas, Leon},
  title     = {Open Semantic Revision Control with {R43ples}},
  booktitle = {Proceedings of the 12th International Conference on
               Semantic Systems (SEMANTiCS)},
  year      = {2016},
  doi       = {10.1145/2993318.2993336}
}

@inproceedings{ibanez2012suset,
  author    = {Ib{\'a}{\~n}ez, Luis-Daniel and Skaf-Molli, Hala and
               Molli, Pascal and Corby, Olivier},
  title     = {Synchronizing Semantic Stores with Commutative Replicated
               Data Types},
  booktitle = {Proceedings of the 21st International Conference on World Wide
               Web ({WWW} '12 Companion)},
  pages     = {1091--1096},
  year      = {2012},
  doi       = {10.1145/2187980.2188246},
  note      = {Defines {SU-Set}. Extended version: Ib{\'a}{\~n}ez et al.,
               Live Linked Data: Synchronising Semantic Stores with Commutative
               Replicated Data Types, \emph{International Journal of Metadata,
               Semantics and Ontologies} 8(2):119--133, 2013,
               \textsc{doi}:10.1504/IJMSO.2013.056605}
}

@misc{terminusdb,
  key          = {TerminusDB},
  title        = {{TerminusDB}: a versioned graph database},
  howpublished = {Software, \url{https://terminusdb.com/}},
  url          = {https://terminusdb.com/},
  year         = {2024}
}

@misc{choudhary2026receiptstrustworthycleanupagentbuilt,
      title={Keep Your Receipts: Trustworthy Cleanup of Agent-Built Knowledge Graphs},
      author={Vishisht Choudhary},
      year={2026},
      howpublished={arXiv:2610.17719},
      eprint={2610.17719},
      archivePrefix={arXiv},
      primaryClass={cs.AI},
      url={https://arxiv.org/abs/2610.17719},
}

@misc{raj2026plantmemorytechnicallanguage,
      title={Plant Memory: Technical Language Processing and a Verified Maintenance Ontology for Agent-Accessible Failure Knowledge from {SAP PM} Work Orders},
      author={Akash Raj},
      year={2026},
      howpublished={arXiv:2610.19131},
      eprint={2610.19131},
      archivePrefix={arXiv},
      primaryClass={cs.AI},
      url={https://arxiv.org/abs/2610.19131},
}

@misc{typesafe2026jev,
  author       = {{TypeSafe AI}},
  title        = {{Jev}: Models and System One},
  howpublished = {Model documentation},
  year         = {2026},
  url          = {https://docs.typesafe.ai/models},
  note         = {Versioned model jev-1.13.0; see also https://docs.typesafe.ai/concepts/system-one. Accessed 22 October 2026}
}

@misc{knublauch2017shacl,
  author       = {Knublauch, Holger and Kontokostas, Dimitris},
  title        = {Shapes Constraint Language ({SHACL})},
  howpublished = {W3C Recommendation, 20 July},
  year         = {2017},
  url          = {https://www.w3.org/TR/2017/REC-shacl-20170720/}
}

\end{document}